\documentstyle[aps]{revtex}
\begin{document}
\draft
\author{L.~Gamberale}
\address{Physics Department, University of Milan}
\author{G.~Preparata}
\address{Physics Department, University of Milan}
\author{She-Sheng Xue}
\address{INFN-section of Milan, Via Celoria 16, Milan, Italy}
\date{\today}
\title{External magnetic fields in gauge theories}
\maketitle
\begin{abstract}
A general discussion is presented of the response of a gauge-field system to
external magnetic fields, in the light of a theorem due to S.~Elitzur. As a
result a natural understanding emerges of some recent puzzling results from
lattice MC simulations, as well as of the phenomenon of ``perfect
diamagnetism''
of non-abelian gauge theories, discovered almost ten years ago.
\end{abstract}

Since the publication of the original work of Savvidy in 1977 \cite{s} a great
deal of attention has been devoted to the understanding of the response of
gauge field systems to external magnetic fields. The reason is obvious:
in this way one hopes to obtain crucial in formation about the
non-perturbative structure of the quantum-mechanical ground state. Today's
theoretical situation can be summarized as follows:
Savvidy's calculation of the effective potential $E(B)$, function
of the magnetic field B, in $SU(2)$ yields up to $O(g^2)$ the result ($\Lambda$
is the ultraviolet cutoff)
\begin{equation}
E(B)={B^2\over2}-{11\over 48\pi^2}g^2B^2\ln\left({\Lambda^2\over gB}\right)
+O\left[g^4B^2\ln\left({\Lambda^2\over gB}\right)\right],
\label{s}
\end{equation}
in agreement with the perturbative renormalization group prediction. Eq.~(\ref
{s}) thus strongly corroborated the universal expectation that the real QCD
ground state would condense an abelian magnetic field $B^*$, and its size
$B^*\sim O(\Lambda^2_{QCD})$ would be in complete agreement with perturbative
QCD (PQCD) at short space-time distances, or equivalently for momenta
$|p_\mu|\geq (B^*)^{1\over 2}.$ In this way a confining mechanism was
identified in the true ground state ($B^*\not=0$), that
{\it would not interfere} with the universally accepted and utilized PQCD.

Unfortunately Savvidy's calculation was found in error by Nielsen and Olesen
\cite{no}, who discovered that Savvidy had neglected the contribution to the
effective potential of a set of modes, {\it the unstable modes}, that in a
constant magnetic field carry a negative squared energy. The effect of such
neglected modes upon the true value of $E(B)$ was finally computed in 1985 by
one of us (G.P.)\cite{gp}, who showed that their inclusion modified
Eq.~(\ref{s}) by substituting the classical energy density term ${B^2\over 2}$
with the much weaker term $k(gB)^2$. With such correction the rosy
(for PQCD) scenario painted by Savvidy changed completely: on one hand one
found a falsification of the Asymptotic Freedom (AF) Ansatz and of its
consequence:
PQCD, on the other an interesting picture emerged of the (probably)
true , color confining QCD ground state \cite{gp2}.

In spite of its non-negligible potentiality for a correct understanding of QCD
these latter conclusions have been essentially ignored by the scientific
community, the only exceptions known to us being a paper published in 1986
\cite{ma}, immediately rebutted \cite{gp3} without further counterdeductions
and
another published in 1988 \cite{con}, where an independent confirmation of the
falsification of PQCD was presented.

We would like to stress that the central and unexpected result of the original
calculation
\cite{gp} is the {\it complete cancellation} of the classical energy density
${B^2\over 2}$ by the unstable modes, neglected by Savvidy.
The ``non-renormalization'' of the negative logarithmic term of eq.~(\ref{s}),
laboriously proved in \cite{gp}, can in fact easily be derived by use of
the energy-momentum tensor trace anomaly \cite{con,gp4}, and is therefore
completely unproblematic. Thus the crucial issue is whether the
{\it complete cancellation}, or the {\it perfect diamagnetism}, are a real,
general feature of any non-abelian gauge theory.

In the last few years there has been a renewed interest on Savvidy's problem
and new calculations
\cite{x,tw,tw1} have appeared of the energy density $E(B)$ of lattice gauge
theory in
an external magnetic field $H$. As the results obtained allegedly appear of
difficult interpretation \cite{tw,tw1}, this paper wishes to be a contribution
to the
clarification of the important physics issues associated to probing gauge
systems with external magnetic fields.

Let us first briefly introduce the problem and its basic physical quantities.
For definiteness' sake, we shall work in the continuum Euclidean space-time,
the lattice transcription being completely straightforward. The effective
action $W(H)$ is defined through
\begin{equation}
e^{-W(H)V_4}=\int[dA]e^{-S(A)-HF},
\label{e}
\end{equation}
where $S(A)=\int d^4x{1\over4}F^a_{\mu\nu}F^{\mu\nu}_a$ ($a=1,2,3,$ we work
without prejudice
in $SU(2)$) is usual gauge action without gauge
fixing, $V_4$ is the 4-dimensional Euclidean volume of the gauge system and the
abelian operator $F$ is defined as
\begin{equation}
F=\int d^4x (\partial_1 A_2^3-\partial_2 A^3_1).
\label{a}
\end{equation}
The abelian magnetic field $B$ is defined as
\begin{equation}
B(H)={\partial W(H)\over\partial H},
\label{b}
\end{equation}
and corresponds to the expectation value of the operator $F$ with the action
$S(A)+HF$. The effective potential $E(B)$, i.e., the energy density of the
gauge system with the latter action, is then defined as the Legendre transform:
\begin{equation}
E(B)=HB-W(H),
\label{en}
\end{equation}
which implies that the external magnetic field $H$ is just:
\begin{equation}
H={\partial E(B)\over\partial B}.
\label{h}
\end{equation}
Thus if a non-trivial minimum of $E(B)\; (B^*=B(H)|_{H=0}\not=0)$ exists, the
gauge system reaches it {\it
spontaneously}, for according to (\ref{h}) this configuration
corresponds to $H=0$. Around such a
minimum one clearly has
\begin{equation}
E(H)-E(0)\geq 0,
\label{n}
\end{equation}
where by $E(H)$ we clearly mean $E[B(H)]$. Suppose however, that one knows that
$E(B)$ has no non-trivial extremal point, i.e., that $H=0$ for $B^*=B(0)=0$
only, in this case one can prove that
\begin{equation}
\Delta E(H)=E(H)-E(0)\leq 0.
\label{d}
\end{equation}
Indeed, from eq.~(\ref{e}) one has
\begin{equation}
W(H)-W(0)=-{1\over V_4}\ln\langle e^{(-HF)}\rangle,
\label{dw}
\end{equation}
where by $\langle\cdot\cdot\cdot\rangle$ we denote the expectation value
with the respect to the distribution $e^{-S(A)}$. By the ``convexity theorem''
\cite{f} that
stipulates that $\langle e^F\rangle\geq e^{\langle F\rangle}$, one easily gets
for any $H$
\begin{equation}
W(H)-W(0)\leq -{1\over V_4}\ln e^{\langle (-HF)\rangle}=0,
\label{el}
\end{equation}
where the last equality is due to the Elitzur theorem
$\langle F\rangle=B^*=0$ \cite{e}. Using eq.~(\ref{en}) we can cast this
inequality in the form
\begin{equation}
B{\partial\Delta E(B)\over \partial B}-\Delta E(B)\leq 0.
\label{m}
\end{equation}
Now if we assume that $\Delta E(B)\geq 0$ for $B>0\; (H>0)$,
from eq.~(\ref{m}) we get
\begin{equation}
{\Delta E(B)\over B}\leq {\Delta E(B_0)\over B_0}\hskip1cm (B>B_0>0).
\label{if}
\end{equation}
In three and four dimensions, on which our attention is concentrated, based on
dimensional considerations we have that $\lim_{B\rightarrow 0}{\Delta
E(B)\over B}
=0$, thus by letting $B_0\rightarrow 0$ in eq.~(\ref{if}) we obtain $\Delta
E(B)\leq 0$, which is in contradiction with our
assumption $\Delta E(B)\geq 0$. Thus the inequalities (\ref{el}) and (\ref{m})
imply eq.~(\ref{d}), which shows that for $H=0$ the energy density is maximum
and not minimum, like it happens in the case when a non-trivial magnetic field
$B$ ``condenses'' spontaneously in the system.

By transcribing, without any fundamental modification, our results for a
lattice, we immediately realize that in a finite lattice {\it without} gauge
fixing, we {\it do} know that no non-trivial zero ($B^*\not=0$) for ${\partial
E(B)\over \partial B}=H=0$ exists. This is just the content of the quoted
powerful theorem due to the Elitzur \cite{e}, which disallows the spontaneous
breaking of a local gauge symmetry. On the other hand, if one fixes the gauge,
as is done in perturbation theory in continuous gauge theories, the Elitzur
theorem does not apply and in fully generality the gauge system will settle in
the non-trivial ($B^*\not=0$) configuration where according
to (\ref{n}) the energy density is minimum.

These results allow us to propose a simple interpretation of the Monte Carlo
data presented in refs.~\cite{tw}. The fact that $\Delta E(H)$ is negative
around $H=0$ for the simulations without gauge fixing is a consequence of the
Elitzur Theorem, while the opposite behaviour when
the gauge is fixed, as computed in Ref.~\cite{tw1}, is again a consequence of
the inapplicability of
the Elitzur theorem. In this context we would like to stress that for abelian
gauge theories the situation is completely different for in this case the
magnetic field $B$ is locally gauge invariant, thus in general
$B^*=\langle F\rangle\not=0$ and inequality (\ref{el}) does not hold. There
appears no ``perfect diamagnetism'', and analytic \cite{sw} and lattice
computations
\cite{x,tw} completely corroborate this observation.

Finally the validity of the Elitzur theorem demands that the phenomenon of
``perfect diamagnetism'',  by which the classical abelian magnetic field $B$
gets screened by quantum fluctuations, be a universal feature of all
non-abelian
gauge theories, for a classical term ${B^2\over 2}$ only perturbatively
corrected
by the quantum fluctuations certainly contradicts the general result
(\ref{m}). We note that this conclusion as well as the consequent inadequacy
of PQCD are in agreement with a recent paper by Patrascioiu and Seiler
\cite{ps}.

% body of paper here

% now the references. delete or change fake bibitem. delete next three
%   lines and directly read in your .bbl file if you use bibtex.

% \begin{figure}
% \caption{}
% \label{}
% \end{figure}

% \begin{table}
% \caption{}
% \label{}
% \end{tabular}
% \end{table}

\end{document}